\documentclass[twocolumn,floatfix,showpacs,superscriptaddress]{revtex4}
\usepackage{graphicx}
\usepackage{amsmath}
\usepackage{amssymb}
\usepackage{bm}

\begin{document}

\title{Zero-voltage conductance peak from weak antilocalization in a Majorana nanowire}

\author{D. I. Pikulin}
\affiliation{Instituut-Lorentz, Universiteit Leiden, P.O. Box 9506, 2300 RA Leiden, The Netherlands}
\author{J. P. Dahlhaus}
\affiliation{Instituut-Lorentz, Universiteit Leiden, P.O. Box 9506, 2300 RA Leiden, The Netherlands}
\author{M. Wimmer}
\affiliation{Instituut-Lorentz, Universiteit Leiden, P.O. Box 9506, 2300 RA Leiden, The Netherlands}
\author{H. Schomerus}
\affiliation{Department of Physics, Lancaster University, Lancaster, LA1 4YB, United Kingdom}
\author{C. W. J. Beenakker}
\affiliation{Instituut-Lorentz, Universiteit Leiden, P.O. Box 9506, 2300 RA Leiden, The Netherlands}
\date{August 2012}
\begin{abstract}
We show that weak antilocalization by disorder competes with resonant Andreev reflection from a Majorana zero-mode to produce a zero-voltage conductance peak of order $e^{2}/h$ in a superconducting nanowire. The phase conjugation needed for quantum interference to survive a disorder average is provided by particle-hole symmetry --- in the absence of time-reversal symmetry and without requiring a topologically nontrivial phase. We identify methods to distinguish the Majorana resonance from the weak antilocalization effect.
\end{abstract}
\pacs{74.45.+c, 73.63.Nm, 74.25.fc, 74.78.Na}
\maketitle

\section{Introduction}
\label{intro}

Weak localization (or antilocalization) is the systematic constructive (or destructive) interference of phase conjugate series of scattering events. In disordered metals it is time-reversal symmetry that provides for phase conjugation of backscattered electrons and protects their interference from averaging out to zero \cite{Ber84,Lee85}. A magnetic field breaks time-reversal symmetry, changing the disorder-averaged conductance by an amount $\delta G$ of order $e^{2}/h$. The sign of $\delta G$ distinguishes weak localization ($\delta G<0$, conductance dip) from weak antilocalization ($\delta G>0$, conductance peak).

Andreev reflection at a superconductor provides an alternative mechanism for phase conjugation due to particle-hole symmetry. No time-reversal symmetry is needed, so weak (anti)localization can coexist with a magnetic field and is only destroyed by a bias voltage \cite{Bro95,Alt96}. The resulting zero-bias anomaly in the conductance of a normal-metal--superconductor (NS) junction is obscured in zero magnetic field by the much larger effects of induced superconductivity, which scale with the number of transverse modes $N$ in the junction. These order $Ne^{2}/h$ effects are suppressed by a magnetic field, only the order $e^{2}/h$ effect from weak (anti)localization remains \cite{Bee97}.

In a superconducting nanowire there is an altogether different origin of zero-bias anomalies in a magnetic field, namely the midgap state that appears at the NS interface following a topological phase transition \cite{Lut10,Ore10,Ali12}. Resonant Andreev reflection from the zero-mode gives a $2e^{2}/h$ conductance peak at zero voltage \cite{Law09}. The first reports \cite{Mou12,Den12,Das12} of this signature of a Majorana fermion are generating much excitement \cite{Wil12}. There is an urgent need to understand the effects of disorder, in order to determine whether it may produce low-lying resonances that obscure the Majorana resonance \cite{Fle10,Kel12,Tew12,Pie12,Pot12}.

These recent developments have motivated us to investigate the interplay of Majorana zero-modes and weak (anti)localization. Earlier studies of weak (anti)localization at an NS junction \cite{Bro95,Alt96,Sle96,Alt97,Rod10} did not consider the possibility of a topologically nontrivial phase with Majorana fermions. Calculations of the local density of states near a zero-mode \cite{Iva02,Ios12,Bag12} address the same physics of midgap quantum interference that we do, but cannot determine the conductance.

This paper consists of two parts: We first give in Sec.\ \ref{analytics} a simple model of a disordered NS interface that allows us to obtain analytical results for $\delta G$ with and without Majorana zero-modes. We then turn in Sec.\ \ref{simulation} to a numerical simulation of a Majorana nanowire and compare the conductance peak due to weak antilocalization (in the topologically trivial phase) with that from a Majorana zero-mode (in the nontrivial phase). The two effects can appear strikingly similar, but in the concluding Sec.\ \ref{discuss} we will discuss several ways in which they may be distinguished.

Before we present our findings, we wish to emphasise that it is not the purpose of this work to diminish the significance of experiments reporting the discovery of Majorana fermions in superconductors. On the contrary, we feel that existing \cite{Mou12,Den12,Das12} and forthcoming experiments will gain in significance if possible alternative mechanisms for zero-voltage conductance peaks in a magnetic field are identified and understood, so that they can be ruled out. Weak antilocalization was so far overlooked as one such mechanism.

\section{Analytical theory}
\label{analytics}

\begin{figure}[tb]
\centerline{\includegraphics[width=0.9\linewidth]{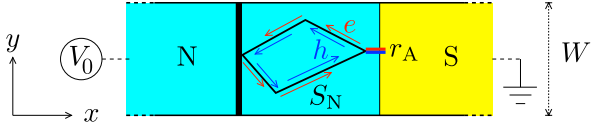}}
\caption{A bias voltage $V_{0}$ applied to the normal metal (N) drives a current $I$ into the grounded superconductor (S). Electrons and holes ($e,h$) are scattered by disorder or a tunnel barrier in N and converted into each other by Andreev reflection at the NS interface, as described by the scattering matrices $S_{\rm N}$ and $r_{\rm A}$. Particle-hole symmetry ensures that the phase shifts accumulated by $e$ and $h$ along a closed trajectory cancel, irrespective of whether time-reversal symmetry is broken or not. Such phase conjugate series of scattering events permit weak (anti)localization to persist in a magnetic field.
}
\label{fig_NS}
\end{figure}

For the analytical calculation we consider a superconducting wire that supports $Q$ topologically protected zero-modes at the interface with a normal metal (see Fig.\ \ref{fig_NS}). The stability of Majorana zero-modes depends crucially on the fundamental symmetries of the system \cite{Ryu10}. At most a single zero-mode is topologically protected if both time-reversal symmetry is broken (by a magnetic field) and spin-rotation symmetry is broken (by spin-orbit coupling), so that only particle-hole symmetry remains. This is called symmetry class D with $Q\in\{0,1\}$. If the wire is sufficiently narrow (relative to the spin-orbit coupling length), an approximate chiral symmetry \cite{Tew11,note0} stabilizes up to $N$ zero-modes. (The integer $N$ is the number of propagating electronic modes through the wire in the normal state, counting both spin and orbital degrees of freedom.) This is called symmetry class BDI with $Q\in\{0,1,2,\ldots N\}$. 

\subsection{Scattering matrix}
\label{sec_Smatrix}

We construct the scattering matrix of the NS junction at the Fermi level by assuming a spatial separation of normal scattering in N and Andreev reflection in S. Within the excitation gap there is no transmission through the superconductor. The matrix $r_{\rm A}$ of Andreev reflection amplitudes from the superconductor is then a $2N\times 2N$ unitary matrix. Mode mixing at the NS interface can be incorporated in the scattering matrix $S_{\rm N}$ of the normal region, so we need not include it in $r_{\rm A}$. It has the block form \cite{Bee11,Die12}
\begin{align}
&r_{\rm A}=\begin{pmatrix}
\Gamma&\Lambda\\
\Lambda^{\ast}&\Gamma
\end{pmatrix},\;\;
\Gamma=\bigoplus_{m=1}^{M}\begin{pmatrix}
\cos\alpha_{m}&0\\
0&\cos\alpha_{m}
\end{pmatrix}\oplus \emptyset_{Q}\oplus\openone_{\zeta},\nonumber\\
&\Lambda=\bigoplus_{m=1}^{M}\begin{pmatrix}
0&-i\sin\alpha_{m}\\
i\sin\alpha_{m}&0
\end{pmatrix}\oplus \openone_{Q}\oplus\emptyset_{\zeta}.\label{rAdef}
\end{align}
We have defined $\zeta =0$ if the difference $N-Q$ is even and $\zeta =1$ if $N-Q$ is odd, so that $N-Q-\zeta\equiv 2M$ is an even integer. The Andreev reflection eigenvalues $\rho_{m}=\sin^{2}\alpha_{m}$ that are not pinned at 0 or 1 are twofold degenerate \cite{Ber09}.

The symbols $\openone_{n},\emptyset_{n}$ denote, respectively, an $n\times n$ unit matrix or null matrix for $n\geq 1$. The empty set is intended for $n=0$. To make the notation more explicit, we give some examples of the direct sums,
\begin{align}
&\openone_{1}\oplus\emptyset_{1}=\begin{pmatrix}
1&0\\
0&0
\end{pmatrix},\;\;\openone_{2}\oplus\emptyset_{1}=\begin{pmatrix}
1&0&0\\
0&1&0\\
0&0&0
\end{pmatrix},\nonumber\\
&\openone_{2}\oplus\emptyset_{0}=\begin{pmatrix}
1&0\\
0&1
\end{pmatrix},\;\;\openone_{1}\oplus\emptyset_{0}=1,\;\;\openone_{0}\oplus\emptyset_{1}=0.
\label{examples}
\end{align}

The normal region has scattering matrix
\begin{equation}
S_{\rm N}=\begin{pmatrix}
s_{0}&0\\
0&s_{0}^{\ast}
\end{pmatrix},\;\;s_{0}=\begin{pmatrix}
r'&t'\\
t&r
\end{pmatrix}.\label{SNdef}
\end{equation}
The electron and hole blocks (with $N\times N$ reflection and transmission matrices $r,r',t,t'$) are each others complex conjugate at the Fermi level. The off-diagonal blocks of $S_{\rm N}$ vanish, because the normal metal cannot mix electrons and holes. The matrix $s_{0}$ is unitary, $s_{0}s_{0}^{\dagger}=1$, without further restrictions in class D. In class BDI chiral symmetry requires that $s_{0}=s_{0}^{\rm T}$ is also a symmetric matrix.

To separate the mixing of modes from backscattering, we make use of the polar decomposition
\begin{equation}
s_{0}=\begin{pmatrix}
U&0\\
0&V
\end{pmatrix}\begin{pmatrix}
-\sqrt{1-{\cal T}}&\sqrt{\cal T}\\
\sqrt{\cal T}&\sqrt{1-{\cal T}}
\end{pmatrix}\begin{pmatrix}
U'&0\\
0&V'
\end{pmatrix}.\label{polardec}
\end{equation}
The matrices $U,V,U',V'$ are $N\times N$ unitary matrices and ${\cal T}={\rm diag}\,(T_{1},T_{2},\ldots T_{N})$ is a diagonal matrix of transmission eigenvalues of the normal region. In class BDI chiral symmetry relates $U'=U^{\rm T}$, $V'=V^{\rm T}$.

\subsection{Conductance}
\label{sec_conductance}

We combine $S_{\rm N}$ and $r_{\rm A}$ to obtain the matrix $r_{he}$ of Andreev reflection amplitudes (from electron $e$ to hole $h$) of the entire system. This calculation is much simplified in the case $\zeta=0$, $\rho_{m}=1$ ($m=1,2,\ldots M$) that all modes at the NS interface are Andreev reflected with unit probability. For this case $\Gamma=0$, $N-Q=2M$, we obtain
\begin{equation}
r_{he}=t'^{\ast}\Lambda^{\ast}(1-r\Lambda r^{\ast}\Lambda^{\ast})^{-1}t,\;\;\Lambda=\sigma_{y}^{\oplus M}\oplus \openone_{Q}.\label{rhedef}
\end{equation}
The notation $\sigma_{y}^{\oplus M}$ signifies the $2M\times 2M$ matrix constructed as the direct sum of $M$ Pauli matrices.

The Andreev reflection matrix determines the conductance
\begin{equation}
G=G_{0}\,{\rm Tr}\,r_{he}^{\vphantom{\dagger}}r_{he}^{\dagger},\;\;G_{0}=2e^{2}/h.\label{Gdef}
\end{equation}
Substitution of the polar decomposition \eqref{polardec} gives the compact expression
\begin{equation}
\begin{split}
&G/G_{0}={\rm Tr}\,{\cal T}{\cal M}{\cal T}{\cal M}^{\dagger},\\
&{\cal M}=(1-\Omega^{\ast}\sqrt{1-{\cal T}}\Omega\sqrt{1-{\cal T}})^{-1}\Omega^{\ast},\;\;\Omega=V'\Lambda V^{\ast}.
\end{split}
\label{Gpolar}
\end{equation}

This is the zero-temperature conductance at the Fermi level, in the limit of zero bias voltage. Away from the Fermi level particle-hole symmetry is broken, so the electron and hole blocks in $S_{\rm N}$ are distinct unitary matrices $s_{e}$ and $s_{h}$. If the bias voltage $V_{0}$ remains small compared to the excitation gap, we can keep the same $r_{\rm A}$. The finite-voltage differential conductance $\tilde{G}=dI/dV_{0}$ is then given by
\begin{equation}
\begin{split}
&\tilde{G}/G_{0}={\rm Tr}\,{\cal T}_{h}\tilde{\cal M}{\cal T}_{e}\tilde{\cal M}^{\dagger},\\
&\tilde{\cal M}=(1-\Omega_{h}^{\ast}\sqrt{1-{\cal T}_{e}}\Omega_{e}\sqrt{1-{\cal T}_{h}})^{-1}\Omega_{h}^{\ast},\\
&\Omega_{e}=V'_{e}\Lambda V_{h}^{\ast},\;\;\Omega_{h}=V'_{h}\Lambda V_{e}^{\ast}.
\end{split}\label{Gpolarhighbias}
\end{equation}
The electron matrices are evaluated at energy $eV_{0}$ above the Fermi level and the hole matrices at energy $-eV_{0}$ below the Fermi level.
Chiral symmetry remains operative away from the Fermi level, hence $V'_{e}=V^{\rm T}_{e}$, $V'_{h}=V^{\rm T}_{h}$ $\Rightarrow$ $\Omega_{h}=\Omega_{e}^{\dagger}$ in class BDI. We will apply Eq.\ \eqref{Gpolarhighbias} to voltages large compared to the Thouless energy, when the electron and hole matrices may be considered to be statistically independent.

\subsection{Random matrix average}
\label{sec_randomaverage}

\begin{figure}[tb]
\centerline{\includegraphics[width=0.9\linewidth]{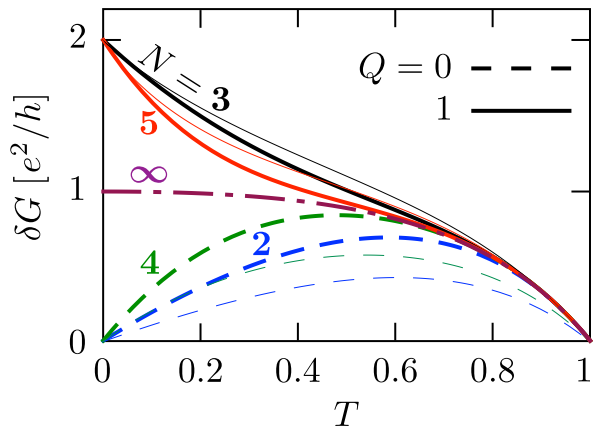}}
\caption{Amplitude $\delta G$ of the average zero-voltage conductance peak as a function of (mode-independent) transmission probability $T$, in symmetry class D (thick curves) and BDI (thin curves) for different number of modes $N$. The superconductor is topologically trivial when $N$ is even ($Q=0$, dashed curves) and nontrivial when $N$ is odd ($Q=1$, solid curves). The dash-dotted curve is the $Q$-independent large-$N$ limit \eqref{GDlargeN}.
}
\label{fig_smallNpeak}
\end{figure}

Isotropic mixing of the modes by scattering in the normal region means that the unitary matrices in the polar decomposition \eqref{polardec} are uniformly distributed in the unitary group ${\cal U}(N)$. We can calculate the average conductance for a given set of transmission eigenvalues by integration over ${\cal U}(N)$ with the uniform (Haar) measure. A full average would then still require an average over the $T_{n}$'s, but if these are dominated by a tunnel barrier they will fluctuate little and the partial average over the unitary matrices is already informative.

The calculation is easiest if all $T_{n}$'s have the same value $0\leq T\leq 1$. The average zero-voltage conductance $\langle G\rangle$ is then given by the integral
\begin{equation}
\langle G\rangle=T^{2}G_{0}\int_{0}^{2\pi}d\phi\,\rho(\phi)\left|1-(1-T)e^{i\phi}\right|^{-2},\label{Gav}
\end{equation}
with $\rho(\phi)=\langle\sum_{n}\delta(\phi-\phi_{n})\rangle$ the density on the unit circle of the eigenvalues $e^{i\phi_{n}}$ of $\Omega\Omega^{\ast}$. The corresponding finite-voltage expression has a uniform $\rho=N/2\pi$, leading to
\begin{equation}
\langle\tilde G\rangle=NG_{0}T/(2-T),\label{Gtildeav}
\end{equation}
irrespective of the symmetry class and independent of the topological quantum number $Q$.

The zero-voltage average \eqref{Gav} does depend on $Q$ and is different for class D and BDI. The calculations are given in the Appendix. Explicit expressions in class D are
\begin{align}
\frac{\langle G\rangle_{\rm D}}{G_{0}}=\begin{cases}
2T&{\rm for}\;\;N=2,\;\;Q=0,\\
1+2T^{2}&{\rm for}\;\;N=3,\;\;Q=1,\\
2T(2-T+T^2)&{\rm for}\;\;N=4,\;\;Q=0,\\
1+2T^{2}(3-2T+T^{2})&{\rm for}\;\;N=5,\;\;Q=1.
\end{cases}\label{GDsmallN}
\end{align}
The $Q$-dependence appears to second order in the reflection probability $R=1-T$, while the general first-order result
\begin{equation}
\langle G/G_{0}\rangle_{\rm D}=N(1-2R)+2R+{\cal O}(R^{2})\label{GDsmallR}
\end{equation}
is $Q$-independent. The corresponding expressions in class BDI are more lengthy, and we only record the small-$R$ result
\begin{equation}
\langle G/G_{0}\rangle_{\rm BDI}=N(1-2R)+2R\frac{Q^{2}+N}{N+1}+{\cal O}(R^{2}),\label{GBDIsmallR}
\end{equation}
to show that it is $Q$-dependent already to first order in $R$. These are all finite-$N$ results. In the large-$N$ limit the $Q$-dependence is lost,
\begin{equation}
\langle G/G_{0}\rangle=\frac{NT}{2-T}+\frac{2(1-T)}{(2-T)^{2}}+{\cal O}(N^{-1}),\label{GDlargeN}
\end{equation}
irrespective of the symmetry class.

As illustrated in Fig.\ \ref{fig_smallNpeak}, for this case that all $T_{n}$'s have the same value $T$ the difference $\delta G=\langle G\rangle-\langle \tilde{G}\rangle$ is positive, corresponding to weak \textit{anti}localization and a conductance \textit{peak}. The sign of the effect may change if the $T_{n}$'s are very different, in particular in class BDI --- which has $\delta G<0$ in a quantum dot geometry (circular ensemble) \cite{Die12}. This is a special feature of quantum interference in a magnetic field, that the distinction between weak localization and antilocalization is not uniquely determined by the symmetry class \cite{Rod10,Whi09,Eng11}.

\section{Simulation of a microscopic model}
\label{simulation}

The random-matrix calculation serves a purpose for a qualitative understanding of the weak antilocalization effect. For a quantitative description we need to relax the assumption of channel-independent $T_{n}$'s. For that purpose we now turn to a microscopic model of a Majorana nanowire.

\subsection{Model Hamiltonian}
\label{sec_modelH}

\begin{figure*}[tb]
\centerline{\includegraphics[width=0.8\linewidth]{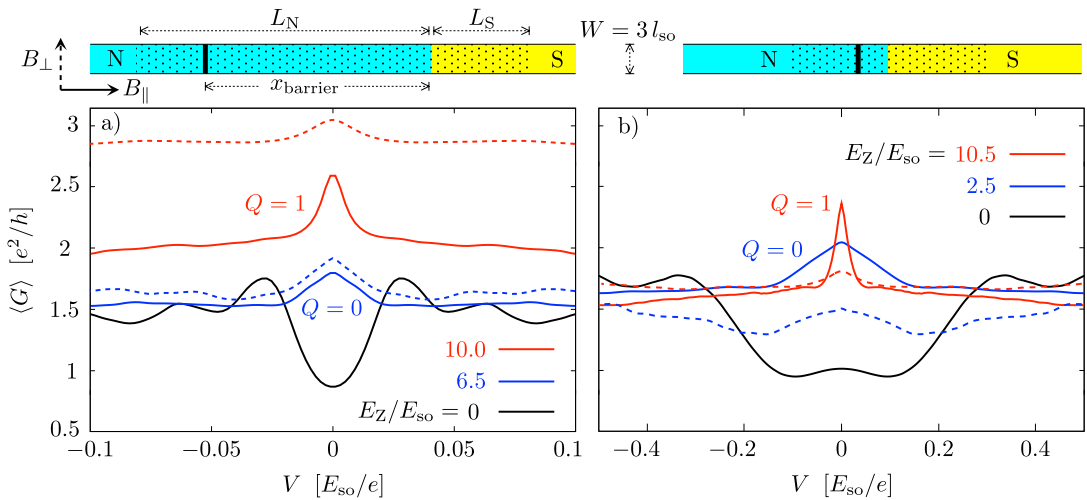}}
\caption{Disorder-averaged differential conductance as a function of bias voltage, for a nanowire modeled by the Hamiltonian \eqref{HM}. The two panels a) and b) correspond to the two geometries shown to scale above each plot. (The solid vertical line indicates the position of the tunnel barrier, relative to the NS interface; disordered regions are dotted.) Each panel shows data for zero magnetic field (black), and for two nonzero magnetic field values (blue and red). The solid curves are for parallel field $B_{\parallel}$ and the dashed curves for perpendicular field $B_{\perp}$. The system is topologically trivial ($Q=0$) in all cases except for the red solid curves ($Q=1$). (The parameter values are listed in Ref.\ \onlinecite{paramfig3}.)
}
\label{fig_classD}
\end{figure*}

Folowing Refs.\ \onlinecite{Lut10,Ore10}, we consider a conducting channel parallel to the $x$-axis on a substrate in the $x$-$y$ plane (width $W$, Fermi energy $E_{\rm F}$), in a magnetic field $\bm{B}$ (orientation $\hat{n}$, Zeeman energy $E_{\rm Z}=\frac{1}{2}g_{\rm eff}\mu_{\rm B} B$), with Rashba spin-orbit coupling (characteristic energy $E_{\rm so}=m_{\rm eff}\alpha_{\rm so}^{2}/\hbar^{2}$, length $l_{\rm so}=\hbar^{2}/m_{\rm eff}\alpha_{\rm so}$), and induced \textit{s}-wave superconductivity (excitation gap $\Delta_{0}$). The Hamiltonian is
\begin{equation}
\begin{split}
&{\cal H}=\begin{pmatrix}
H_{0}-E_{\rm F}&\Delta\sigma_{y}\\
\Delta^{\ast}\sigma_{y}&E_{\rm F}-H_{0}^{\ast}
\end{pmatrix},\\
&H_{0}=\frac{p_{x}^{2}+p_{y}^{2}}{2m_{\rm eff}}+U(x,y)+\frac{\alpha_{\rm so}}{\hbar}(\sigma_{x}p_{y}-\sigma_{y}p_{x})+E_{\rm Z}\hat{n}\cdot\bm{\sigma}.
\end{split}\label{HM}
\end{equation}

The electrostatic potential $U=U_{\rm gate}+\delta U$ contains the gate potential $U_{\rm gate}$ that creates the tunnel barrier and the impurity potential $\delta U$ that varies randomly from site to site on a square lattice (lattice constant $a$), distributed uniformly in the interval $(-U_{\rm disorder},U_{\rm disorder})$. The disordered region is $-L_{\rm N}<x<L_{\rm S}$, an NS interface is constructed by increasing the pair potential $\Delta$ from $0$ to $\Delta_{0}$ at $x=0$, and a rectangular barrier of height $U_{\rm barrier}$, thickness $\delta L_{\rm barrier}$, is placed at $x=-x_{\rm barrier}$. The conductance of the normal region ($x<0$) contains a contribution $G_{\rm disorder}$ from disorder and $G_{\rm barrier}$ from the barrier.

The orientation of the magnetic field plays an important role \cite{Lut10,Ore10}: It lies in the $x$-$y$ plane to eliminate orbital effects on the superconductor and we will only include its effect on the electron spin (through the Zeeman energy). A topologically nontrivial phase needs a nonzero excitation gap for $E_{\rm Z}>\Delta_{0}$, which requires a parallel magnetic field $B_{\parallel}$ ($\hat{n}=\hat{x}$). We will consider that case in the next subsection, and then discuss the case of a perpendicular magnetic field $B_{\perp}$ ($\hat{n}=\hat{y})$ in Sec.\ \ref{sec_fieldrotation}.

\subsection{Average vs.\ sample-specific conductance}
\label{sec_averageG}

\begin{figure*}[tb]
\centerline{\includegraphics[width=0.8\linewidth]{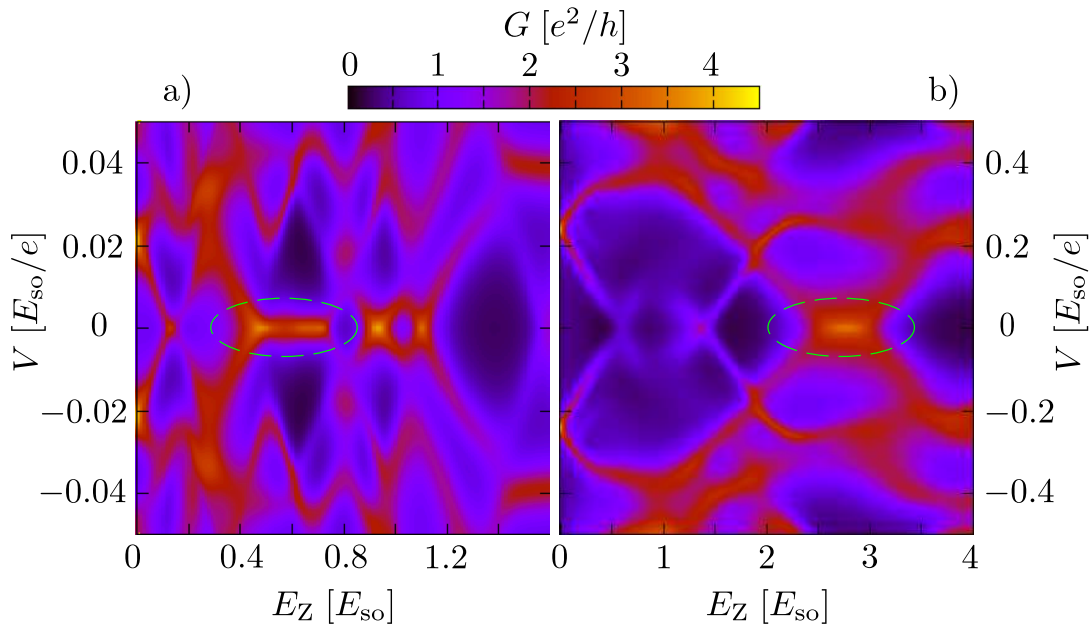}}
\caption{Numerical simulation of a nanowire for a single disorder realization (no averaging). The color scale gives the differential conductance as a function of bias voltage (vertical axis) and parallel magnetic field (horizontal axis). The parameters in panels a,b correspond to those in Fig.\ \ref{fig_classD}a,b, as listed in Ref.\ \onlinecite{paramfig3}. The magnetic field range in both panels is in the topologically trivial phase ($Q=0$), but still exhibits a conductance peak pinned to zero voltage (green circle).
}
\label{fig_Dsingle}
\end{figure*}

To avoid the complications from chiral symmetry we first focus on a relatively wide junction, $W=3\,l_{\rm so}$, when symmetry class D (rather than BDI) applies \cite{Die12}. (We turn to class BDI in the next subsection.) The normal region has $N=8$ propagating modes (including spin) in zero magnetic field, for $E_{\rm F}=12\,E_{\rm so}$. The topological quantum number $Q$ was determined both from the determinant of the reflection matrix \cite{Akh11,note2}, and independently by counting the gap closings and reopenings upon increasing the magnetic field. A transition from $Q=0$ to $Q=1$ is realized by increasing $B_{\parallel}$ at fixed $\Delta_{0}=8\,E_{\rm so}$.

Results are shown in Fig.\ \ref{fig_classD} (solid curves) for two geometries, one with the tunnel barrier far from the NS and another with the barrier close to the interface \cite{paramfig3}.

The disorder-averaged conductance shows a zero-voltage peak in a magnetic field, regardless of whether the nanowire is topologically trivial ($Q=0$) or nontrivial ($Q=1$). The peak disappears in zero magnetic field and instead a conductance minimum develops, indicative of an induced superconducting minigap in the normal region. The two geometries in panels \ref{fig_classD}a and \ref{fig_classD}b show comparable results, the main difference being a broadening of the zero-bias peak when the tunnel barrier is brought closer to the NS interface --- as expected from the increase in Thouless energy \cite{note1}. The shallow maximum which develops around zero voltage in the $B=0$ curve of panel \ref{fig_classD}b is a precursor of the reflectionless tunneling peak, which appears in full strength when the barrier is placed at the NS interface \cite{Bee97}.

This all applies to the average conductance in an ensemble of disordered nanowires. Individual members of the ensemble show mesoscopic, sample-specific conductance fluctuations, in addition to the systematic weak antilocalization effect. For some disorder realizations the zero-voltage conductance peak remains clearly visible, see Fig.\ \ref{fig_Dsingle}. The peak sticks to zero bias voltage over a relatively wide magnetic field range, even though the superconductor is topologically trivial ($Q=0$). The appearance and disappearance of the peak is not associated with the closing and reopening of an excitation gap, so it cannot produce Majorana fermions \cite{Pik11}.

\subsection{Parallel vs.\ perpendicular magnetic field}
\label{sec_fieldrotation}

\begin{figure}[tb]
\centerline{\includegraphics[width=0.8\linewidth]{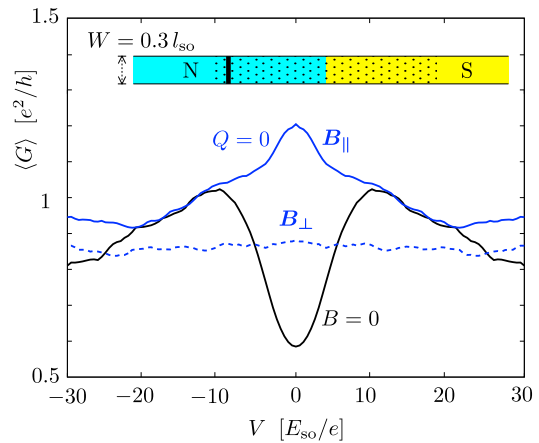}}
\caption{Same as Fig.\ \ref{fig_classD}, but now for a narrower wire in symmetry class BDI (rather than D). The system is topologically trivial, without Majorana zero-modes. The weak antilocalization peak vanishes if the magnetic field is rotated from $B_{\parallel}$ to $B_{\perp}$. (The parameter values are listed in Ref.\ \onlinecite{note3}.)
}
\label{fig_BDIav}
\end{figure}

So far we considered a class D nanowire with magnetic field $B_{\parallel}$ parallel to the wire axis. In a perpendicular magnetic field $B_{\perp}$ (perpendicular to the wire in the plane of the substrate) the symmetry class remains D (broken time-reversal and spin-rotation symmetry), although the topologically nontrivial phase disappears \cite{Lut10,Ore10}. We therefore expect the class D zero-bias peak to persist in a perpendicular field as a result of the weak antilocalization effect. 

This expectation is borne out by the computer simulations, see the dashed curves in Fig.\ \ref{fig_classD}. A zero-bias peak exists for both $B_{\perp}$ and $B_{\parallel}$. If the nanowire is topologically trivial, there is not much difference in the peak height for the two magnetic field directions (compare blue solid and dashed curves). In contrast, if the nanowire is topologically nontrivial for parallel field then the peak is much reduced in perpendicular field (red solid versus dashed curves). The disappearance of the Majorana zero-mode and the collapse of the zero-bias peak may be accompanied by the appearance of propagating modes in the superconducting part of the nanowire. This explains the increased background conductance in the red dashed curve of Fig.\ \ref{fig_classD}a.

The effect of a magnetic field rotation is entirely different when $W\lesssim l_{\rm so}$ and the symmetry class is BDI rather than D \cite{Tew11,Die12}. The term $\sigma_x p_y$ in the Hamiltonian \eqref{HM} can then be neglected, so that ${\cal H}$ commutes with $\sigma_y$ in a perpendicular magnetic field ($\hat{n}=\hat{y}$). The two spin components along $\pm\hat{y}$ decouple and for each spin component separately the particle-hole symmetry is broken. We therefore expect both the Majorana resonance and the weak antilocalization peak to disappear in a perpendicular magnetic field for sufficiently narrow wires.

This is demonstrated by the computer simulations shown in Fig.\ \ref{fig_BDIav}, for the average conductance in a topologically trivial wire of width $W=0.3\,l_{\rm so}$. The main difference with the data in Fig.\ \ref{fig_classD} is that the symmetry class is now BDI rather than D, because of the narrower wire. This change of symmetry class does not significantly affect the weak antilocalization peak in a parallel magnetic field. But if the magnetic field is rotated to a perpendicular direction, the peak disappears --- as expected for a class BDI nanowire.

\subsection{Effects of thermal averaging}
\label{thermalaverage}

\begin{figure}[tb]
\centerline{\includegraphics[width=0.8\linewidth]{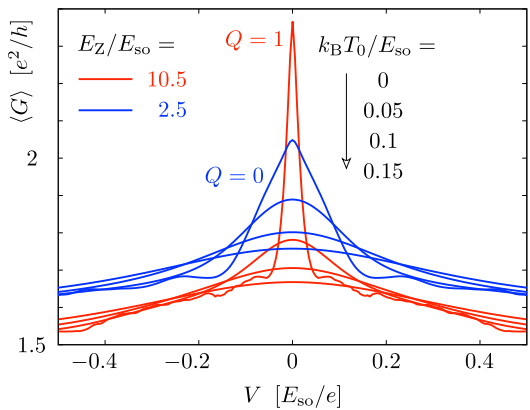}}
\caption{Temperature dependence of the conductance peaks from Fig.\ \ref{fig_classD}b. The four blue curves ($Q=0$, topologically trivial) correspond from top to bottom to four increasing temperatures, and likewise the four red curves ($Q=1$, topologically nontrivial). 
}
\label{fig_finiteT}
\end{figure}

All results presented so far are in the zero-temperature limit. We calculate the temperature dependence of the differential conductance from the finite-$T_{0}$ and finite-$V_{0}$ generalization of Eq.\ \eqref{Gdef},
\begin{align}
&G=\frac{2e}{h}\,\int_{-\infty}^{\infty}d\varepsilon\,\frac{df(\varepsilon-eV_{0})}{dV_{0}}\,{\rm Tr}\,r_{he}^{\vphantom{\dagger}}(\varepsilon)r_{he}^{\dagger}(\varepsilon),\label{GTV}\\
&f(\varepsilon)=\frac{1}{1+\exp(\varepsilon/k_{\rm B}T_{0})}.\label{fdef}
\end{align}
Thermal averaging at a nonzero temperature $T_{0}$ broadens the conductance peak around $V_{0}=0$ and reduces its height, at constant area $\int G\,dV_{0}$ under the peak.

This effect of thermal averaging applies to both the weak antilocalization peak and to the Majorana resonance, but the characteristic temperature scale is different, as shown in Fig.\ \ref{fig_finiteT}. The Majorana zero-mode is more sensitive to thermal averaging because it is more tightly bound to the NS interface, with a smaller Thouless energy and therefore a smaller characteristic temperature.

\section{Discussion}
\label{discuss}

\begin{figure*}[tb]
\centerline{\includegraphics[width=0.8\linewidth]{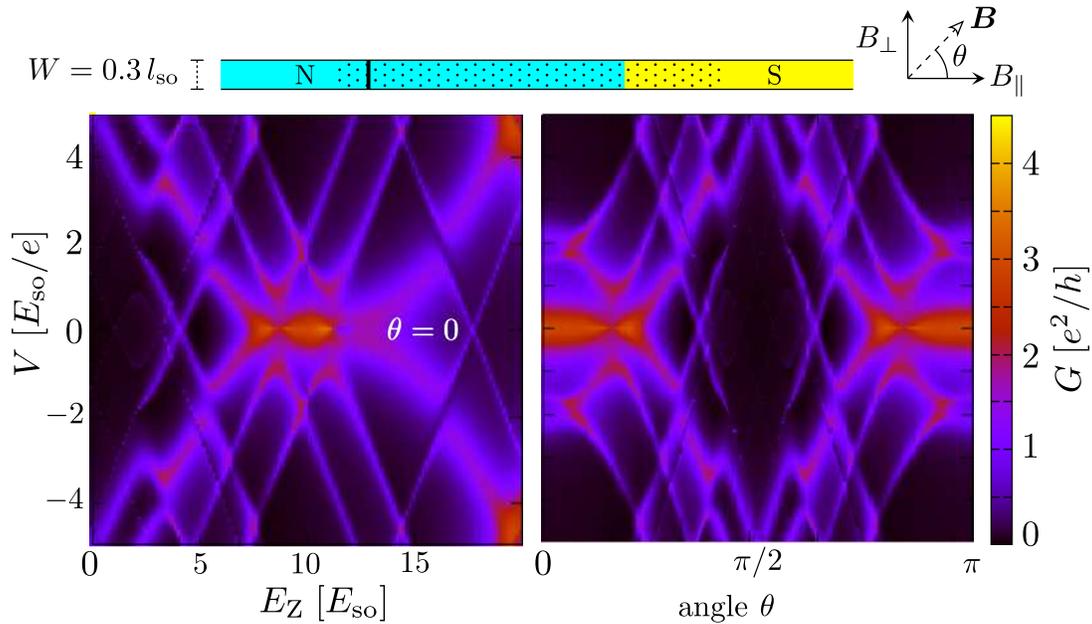}}
\caption{Differential conductance for a single disorder realization of a nanowire ($N=2$ spin-resolved modes, parameter values are listed in Ref.\ \onlinecite{note4}). The left panel shows the appearance of a zero-voltage peak in a range of magnetic field values, for $\bm{B}$ parallel to the wire. The right panel shows the dependence on the orientation of the magnetic field, for a fixed field strength ($E_{\rm Z}=10\,E_{\rm so}$). The zero-voltage peak vanishes if $\bm{B}$ is perpendicular to the wire. This is the same phenomenology as for a Majorana resonance, but here it happens in the topologically trivial phase.
}
\label{fig_BDIsingle}
\end{figure*}

In conclusion, we have shown that random quantum interference by disorder in a superconducting nanowire can systematically produce a zero-voltage conductance peak in the absence of time-reversal symmetry. This weak antilocalization effect relies on the same particle-hole symmetry that protects the Majorana zero-mode, but it exists in both the topologically trivial and nontrivial phase of the superconductor. A conclusive demonstration of Majorana fermions will need to rule out this alternative mechanism for a conductance peak.

There are several strategies one might follow for this purpose:
\begin{itemize}
\item Increasing the tunnel barrier with a gate voltage suppresses the weak antilocalization effect, but not the Majorana resonance. The resonance does become narrower, so at finite temperatures thermal smearing will still lead to a suppression with increasing barrier height and this might not be the most effective strategy to distinguish the two effects.
\item The disappearance of the conductance peak when the magnetic field is rotated (in the plane of the substrate) towards a direction perpendicular to the wire, the technique used in Refs.\ \onlinecite{Mou12,Das12}, can identify the Majorana zero-mode --- but only if the ratio $W/l_{\rm so}$ is sufficiently large that the wire is in class D rather than BDI. In class BDI the Zeeman energy in the rotated field commutes with the Rashba energy, precluding the weak antilocalization effect as well as the Majorana resonance. Both Refs.\ \onlinecite{Mou12,Das12} have $W\lesssim l_{\rm so}$ and are believed to be in class BDI \cite{Tew12,Tew11}, so this complication seems quite relevant.
\item Measuring the conductance through a single-mode point contact is a very effective strategy: for $N=1$ the zero-temperature conductance $G=Q\times 2e^{2}/h$ directly measures the topological quantum number even without any tunnel barrier \cite{Wim11}, and this signature of a Majorana zero-mode is quite robust against finite temperatures. (The chararacteristic energy scale is the induced superconducting gap in the region between the point contact and the superconductor.) The single mode in the point contact should be spin resolved for this to work: If instead the point contact transmits both spins in one orbital mode ($N=2$), then the ambiguity between weak antilocalization and the Majorana resonance remains (see Fig.\ \ref{fig_BDIsingle}).
\item The Majorana resonance from a wire of finite length should split into two at the lowest temperatures, because of the nonzero overlap of the zero-modes at the two ends of the wire \cite{Das12}. No such systematic splitting will occur for the weak antilocalization peak.
\end{itemize}

\acknowledgments

We have benefited from discussions with A. R. Akhmerov and Yu.\ V. Nazarov. The numerical simulations of the nanowire were performed with the {\sc kwant} software package, developed by A. R. Akhmerov, C. W. Groth, X. Waintal, and M. Wimmer. Our research was supported by the Dutch Science Foundation NWO/FOM and by an ERC Advanced Investigator grant.

\appendix

\section{Random-matrix theory}
\label{RMT}

To evaluate the average conductance \eqref{Gav} we seek the density of the eigenvalues $x_{n}=e^{i\phi_{n}}$ of the product $X=\Omega\Omega^{\ast}$ of the unitary matrix $\Omega$ and its complex conjugate. We denote $\mu_n=\cos\phi_n\in[-1,1]$ and determine the joint probability distribution $P(\{\mu_n\})$ using methods from random-matrix theory \cite{For10}.

In symmetry class D, we have $\Omega=V'\Lambda V^{\ast}$ with $V$ and $V'$ independently and uniformly distributed according to the Haar measure $dU$ of the unitary group ${\cal U}(N)$. Because $d(U U')=dU$ for a fixed unitary matrix $U'$, the matrix $\Omega\equiv U$ is itself uniformly distributed in ${\cal U}(N)$.

In class BDI, we have $V'=V^{\rm T}$ and we may write $\Omega\equiv U\lambda U^{\dagger}$ with $U$ uniformly in ${\cal U}(N)$. The diagonal matrix $\lambda={\rm diag}(\lambda_{1},\lambda_{2},\ldots\lambda_{N})$ contains the eigenvalues $\lambda_{n}=\pm 1$ of $\Lambda$. The number $q=|Q|$ of Majorana zero-modes is encoded in the topological invariant  $Q={\rm
Tr}\,\Lambda=\sum_{n}\lambda_{n}$. (For full generality we allow $Q$ to also take on negative values, but the final result will only depend on the absolute value $q$.)

\subsection{Brownian motion of unitary matrices}

We employ Dyson's Brownian motion approach \cite{Dys62}, which sets up a stochastic
process for the unitary matrix $U$ whose stationary distribution coincides with the Haar
measure on ${\cal U}(N)$. In each infinitesimal step of the process, $U\to U\exp(i H)$,
where $H$ is a Hermitian matrix from the Gaussian unitary ensemble, with
identically normal distributed complex numbers $H_{lm}=H_{ml}^*$ ($l\leq
m$), $\overline{H_{lm}} =0$, $\overline{ H_{kl}H_{mn}}
=\delta_{kn}\delta_{lm}\tau$, $\overline{ H_{kl}H_{mn}^*}
=\delta_{km}\delta_{ln}\tau$; the limit $\tau\to 0$ is implied to
generate  infinitesimal increments. 

The corresponding increments $\delta
\mu_n$ can be calculated in perturbation theory. The drift coefficients
$c_l=\lim_{\tau\to 0}\tau^{-1} \overline{\delta \mu_l}$ and the
diffusion coefficients $c_{lm}=\lim_{\tau\to 0}\tau^{-1}
\overline{\delta \mu_l\delta \mu_m}$ follow by averaging over the random
variables in $H$. As we will see, the symmetries in the classes D and BDI
are restrictive enough so that these coefficients can be expressed in
terms of the quantities $\mu_n$ alone, without requiring data from the
eigenvectors of $X$. Thus, the stochastic process for these quantities
closes.

Introducing a fictitious time $t$,  the evolution of the joint
probability distribution is governed by a Fokker-Planck equation,
\begin{equation}
\label{fp}
\frac{\partial P}{\partial t}=\left[-\sum_l\frac{\partial}{\partial \mu_l} c_l+\frac{1}{2}\sum_{l,m}\frac{\partial}{\partial \mu_l}\frac{\partial}{\partial \mu_m}c_{lm}\right]P(\{\mu_n\},t).
\end{equation}
The stationary solution $P(\{\mu_n\})$, for which the right-hand-side of the Fokker-Planck equation vanishes, is the required eigenvalue distribution.

\subsection{Symmetry class D}

In class D we have $X=UU^*$ with $U$ uniformly distributed in ${\cal U}(N)$. Notice that the operation of complex conjugation is basis dependent; if
$B=A^*$ in one basis then this relation is only preserved under
orthogonal transformations, but not under general unitary
transformations.  Thus, we work in a fixed basis $|r\rangle$ (at most
permitting orthogonal basis changes), and define for any
$|\psi\rangle=\sum_r\psi_r|r\rangle$ a complex-conjugated vector
$|\psi^*\rangle\equiv\sum_r\psi_r^*|r\rangle$. As usual,
$\langle\psi|=\sum_r\psi_r^*\langle r|$; thus
$\langle\psi^*|=\sum_r\psi_r\langle r|$.

The matrices $X$ and $U$ are unitary and obey
$\mathrm{Det}\,X=|\mathrm{Det}\,U|^2=1$. Moreover, the matrix $X^*$ has
the same eigenvalues $x_1,x_2,\ldots x_N$ as the matrix $X$. For even
$N$, it follows that all eigenvalues appear in
complex-conjugated pairs; every eigenvalue $x_k$ has a partner $x_{\bar
k}= x^{*}_{k}=x^{-1}_{k}$. For odd $N$, in addition to such pairs there
is a single unpaired eigenvalue, denoted as $x_N$, which (because of the
constraint on the determinant) is pinned at $x_N=1$. The paired
eigenvectors are related according to
\begin{equation}
|\bar k\rangle=\xi_{k}U|k^*\rangle.
\end{equation}
Here we have to set $\xi_{k}$ such that $\xi_{k}^2=\lambda_k$; this
guarantees that the relation between both eigenvectors in a pair is
reciprocal, $|\bar{\bar k}\rangle=|k\rangle$. Observing that the eigenvectors form an
orthogonal basis,  we find the matrix elements
\begin{equation}
\langle k |U | l^*\rangle=\xi_k \delta_{k \bar l}
=(\langle k^* | U^* | l\rangle)^*=\langle l | U^{\rm T} | k^*\rangle.
\label{orthrel}
\end{equation}

With help of these matrix elements we can now evaluate the drift and
diffusion coefficients. In second-order perturbation theory,
\begin{equation}
\delta x_l= \langle l |\delta X| l\rangle+{\sum_k}'\frac{\langle l | \delta X | k\rangle\langle k|\delta X | l \rangle}{x_l-x_k}
,
\end{equation}
where the prime restricts the sum to $k\neq l$ while
\begin{equation}
\delta X= iUHU^*-iXH^*+UHU^*H^* -\tfrac{1}{2}UH^2U^*-\tfrac{1}{2}X{H^*}^2
\end{equation}
is the increment of $X$ to leading order in $\tau$. The Gaussian
averages are now carried out according to the rules
\begin{align}
&\langle k | A H B |l \rangle\langle m | C H D |n \rangle=
\tau
\langle k | A D |n \rangle\langle m | C B |l \rangle,\\
&\langle k | A H B |l \rangle\langle m | C H^* D |n \rangle=
\tau
\langle k | A C^{\rm T} | m^* \rangle\langle n^* | D^{\rm T} B |l \rangle.
\end{align}
In particular,  $\overline{H^2}= N\tau$, $\overline{UHU^*H^*}=\tau
UU^\dagger=\tau$, and
\begin{align}
&\overline{\langle l | UHU^*-XH^* | k\rangle\langle k| UHU^*-XH^* | l \rangle}\nonumber\\
&\quad\quad\quad=2\tau\langle l |X| l \rangle \langle k |X| k \rangle -\tau\langle l | U^T | k^*\rangle   \langle l^* |U^*| k\rangle\nonumber\\
&\quad\quad\quad\quad\quad
-\tau\langle k | U^T | l^*\rangle   \langle k^* |U^*| l\rangle
\nonumber \\
&\quad\quad\quad=
2\tau x_l x_k
-\tau \delta_{l\bar k}( x_l+ x_{\bar l})
,
\end{align}
where we invoked Eq.\ \eqref{orthrel}. We thus obtain
\begin{equation}
\overline{\delta x_l}=
\tau -N\tau x_l
-\tau
{\sum_k}'\frac{2 x_lx_k-\delta_{l\bar k}(x_l+x_{\bar l})
}{x_l-x_k}
.
\end{equation}
Analogously, we find
\begin{equation}
\overline{\delta x_l\delta x_m}
= \overline{\langle l |\delta X| l\rangle\langle m |\delta X| m\rangle}
=-2\tau\delta_{lm}x_l^2+2\tau\delta_{l\bar m}.
\end{equation}
Note that these expressions only depend on the eigenvalues. We remark
that for the pinned unpaired eigenvalue $x_N=1$, occurring if $N$ odd, these
relations deliver $\overline{\delta x_N} =\overline{(\delta x_N)^2}=0$.

We now pass over to the quantities $\mu_l=(x_l+x_{\bar l})/2$, and
restrict the index $l$ such that it enumerates the pairs of eigenvalues.
For even $N$ we then find
\begin{equation}
\overline{\delta \mu_l}
=\tau -2\tau \mu_l
-2\tau(\mu_l^2-1)
{\sum_k}'\frac{1}{\mu_l-\mu_k},\label{incrementDa}
\end{equation}
while for odd  $N$ we have
\begin{equation}
 \overline{\delta \mu_l}
=-3\tau \mu_l
-2\tau(\mu_l^2-1)
{\sum_k}''\frac{1
}{\mu_l-\mu_k},\label{incrementDb}
\end{equation}
where  the double-prime excludes the pinned eigenvalue. Furthermore,
\begin{equation}
\overline{\delta \mu_l\delta \mu_m}
=2\tau(1-\mu_l^2)\delta_{lm}.\label{incrementDc}
\end{equation}

The stationarity condition of the associated Fokker-Planck equation
\eqref{fp} can be expressed as
\begin{equation}\frac{\partial}{\partial \mu_l}\overline{\delta
\mu_l}P=\frac{1}{2}\frac{\partial^2}{\partial \mu_l}\overline{(\delta \mu_l)^2}P.
\label{stationary}
\end{equation}
For even $N=2M$, this is solved by
\begin{subequations}
\label{PmuD}
\begin{equation}
P(\mu_{1},\mu_{2},\ldots\mu_{M})\propto\prod_{k=1}^{M}\frac{1+\mu_k}{\sqrt{1-\mu_k^2}}\,\prod_{l<m=1}^{M}(\mu_l-\mu_m)^2
,\label{PmuDa}
\end{equation}
up to a normalization constant. Each of the $\mu_{n}$'s ($n=1,2,\ldots M$) is twofold degenerate. For odd $N=2M+1$ one eigenvalue is pinned at $+1$, and the remaining ones are twofold degenerate with distribution
\begin{equation}
P(\mu_{1},\mu_{2},\ldots\mu_{M})\propto\prod_{k=1}^{M}\sqrt{1-\mu_k^2}\,\prod_{l<m=1}^{M}(\mu_l-\mu_m)^2.\label{PmuDb}
\end{equation}
\end{subequations}

This concludes our derivation of the eigenvalue distribution of $UU^{\ast}$ with $U$ uniform in ${\cal U}(N)$. We have not found the result \eqref{PmuD} in the literature, but there is a curious correspondence with the known \cite{For10,Gir85} eigenvalue distribution of orthogonal matrices (uniformly distributed according to the Haar measure). An $(N+1)\times(N+1)$ orthogonal matrix $O$ with determinant $-1$ has one eigenvalue pinned at $-1$. If we exclude that eigenvalue, the remaining $N$ eigenvalues of $O$ have same probability distribution as the $N$ eigenvalues of $UU^{\ast}$.   

\subsection{Brownian motion of orthogonal matrices}

As an independent demonstration of this correspondence between the eigenvalue distributions of $UU^{\ast}$ and $O$, we have investigated the Brownian motion of orthogonal matrices. Let $O$ be a random $(N+1)\times(N+1)$-dimensional matrix in the orthogonal group, constrained to the sector ${\rm Det}\,O=-1$.

The Brownian motion is induced by $O(1+A+A^2/2)$, where (in the fixed basis) $A=-A^{\rm T}$ is a real antisymmetric matrix, with $\overline{A_{lm}^2}=\tau$.
Due to the condition on the determinant, there is always one eigenvalue pinned at $x_{N+1}=-1$, while an additional eigenvalue
is pinned at $x_{N}=1$ if $N$ is odd. All other eigenvalues appear in pairs $x_l$, $x_{\bar l}$, with $|\bar l\rangle=| l^*\rangle$ (no additional factors are required).

We calcaluate the increments and average:
\begin{align}
&\delta x_l=\tfrac{1}{2}\langle l |O A^2| l \rangle +\sum_{k\neq l}\frac{\langle l |O A| k \rangle\langle k |O A| l \rangle}{x_l-x_k}\nonumber\\
&\Rightarrow \overline{\delta x_l}=-\tfrac{1}{2}\tau Nx_l+\tau\sum_{k\neq l}\frac{x_l x_k(\delta_{k\bar l}-1)}{x_l-x_k},\\
&\delta x_l\delta x_k=\langle l |O A| l \rangle\langle k |O A| k \rangle\nonumber\\
&\Rightarrow \overline{\delta x_l\delta x_k}=\tau x_l x_k  (\delta_{l\bar k}-\delta_{lk})=\tau   (\delta_{l\bar k}-x_l^2\delta_{lk}).
\end{align}
(Note that $\langle l |A| l \rangle$ does not vanish if $|l\rangle$ is complex, as is generally the case for the unpinned eigenvalues.)

As before, in passing over to $\mu_l$ we restrict indices to enumerate different pairs.
For $N$ even, we find [considering that the restricted sum has $(N-2)/2$ terms]
\begin{equation}
\begin{split}
\overline{\delta \mu_l}={}&\tfrac{1}{2}\tau-\tfrac{1}{2}\tau N\mu_l-
\tau\sum_{k\neq l,N+1}\frac{\mu_l\mu_k-1}{\mu_l-\mu_k}\\
={}&\tfrac{1}{2}\tau-\tau\mu_l- \tau(\mu_l^2-1)\sum_{k\neq l,N+1}\frac{1}{\mu_l-\mu_k},
\end{split}\label{incrementOa}
\end{equation}
while if $N$ is odd [where the restricted sum has $(N-3)/2$ terms],
\begin{equation}
\begin{split}
\overline{\delta \mu_l}={}&-\tfrac{1}{2}\tau N\mu_l-
\tau\sum_{k\neq l,N,N+1}\frac{\mu_l\mu_k-1}{\mu_l-\mu_k}\\
={}&-\tfrac{3}{2}\tau\mu_l-
\tau(\mu_l^2-1)\sum_{k\neq l,N,N+1}\frac{1}{\mu_l-\mu_k}.
\end{split}\label{incrementOb}
\end{equation}
Furthermore,
\begin{equation}
\overline{\delta \mu_l\delta \mu_k}=\tau(1-\mu_l^2)\delta_{lk}.\label{incrementOc}
\end{equation}

Comparison with Eqs.\ \eqref{incrementDa}--\eqref{incrementDc} shows that these are the same average increments, if we rescale $\tau$ by a factor 2. The eigenvalues of $UU^{\ast}$ and $O$ therefore execute the same Brownian motion process, with the same stationary solution \eqref{PmuD}.

\subsection{Symmetry class BDI}

In  class BDI we have $X=U\lambda
U^\dagger U^* \lambda U^{\rm T}$, with $U$ uniform in ${\cal U}(N)$ and $\lambda$ a fixed diagonal matrix with entries $\pm
1$ that sum up to $Q$.
Since here the matrix $X$ is symmetric, $X=X^{\rm T}$, it is now diagonalized
by an orthogonal transformation; thus, the eigenvectors
$|k\rangle=|k^*\rangle$ are real. As in class D, eigenvalues appear
in complex-conjugate pairs, apart from eigenvalues pinned at $1$. We
observe that $\Omega$ mediates between the associated eigenvector, $|\bar
k\rangle=\xi_k\Omega|k\rangle =\xi_k^*\Omega^*|k\rangle$. In order
to treat the partners symmetrically we have to require that that $|\bar
k\rangle$ is also real, so $\xi_k$ compensates any complex
overall factor. It then follows that $\langle k|\Omega\Omega^*|k\rangle=
\xi_k^{2}=\lambda_k$, and thus the coefficients $\xi_k$ are related
to the eigenvalues as in class D.

To identify the pinned eigenvalues note that
$\Omega=\Omega^\dagger=\Omega^{-1}$ is both Hermitian and unitary, and thus has eigenvalues $\pm 1$. Let $\Omega_\pm$ be the
eigenspace for each set of eigenvalues, and  $\Omega^*_\pm$ the analogous
eigenspace for $\Omega^*$, which is spanned by the complex-conjugated
vectors. We denote $\xi=\mathrm{sign}\, Q$. The space
$[\mathrm{span}(\Omega_{-\xi},\Omega_{-\xi}^*)]^\bot$ is then of
dimension $q=|Q|$ (barring accidental degeneracies), and all of the
vectors in this space obey $X|k\rangle=|k\rangle$. Thus $X$ has $q=|Q|$
eigenvalues pinned at 1. For each pinned eigenvalue, insisting that
$|\bar k\rangle=|k\rangle$ implies
$\Omega|k\rangle=\Omega^*|k\rangle=\xi |k\rangle$,
$\xi=\mathrm{sign}\,Q=\pm 1$ (consistent with the property that these
states lie in the joint subspace of $\Omega_{\xi}$ and
$\Omega_{\xi}^*$).

With these additional properties in hand, the evaluation of drift and
diffusion coefficients can proceed along the same steps as before. With
the specified form of $X$, the incremental step of $U$ carries over to an
increment
\begin{align}
\delta X={}&
iU[H,\lambda]U^\dagger \Omega^*
-i\Omega U^*[H^*,\lambda]U^{\rm T}
\nonumber
\\
&+\tau Q (\Omega^*+\Omega)
+2\tau(1-X)-2N\tau X,
\end{align}
where we already averaged terms of second order in $H$; in particular,
terms such as $\overline{UH\lambda HU^\dagger U^*\lambda U^{\rm T}}=\tau Q
\Omega^*$ produce the topological invariant $Q$.
The associated eigenvalue increment averages to
\begin{widetext}
\begin{align}
\overline{\delta x_l}={}& -2N\tau x_l
+\tau Q \langle l |\Omega^*+\Omega|l\rangle
+2\tau(1-x_l)
\nonumber \\
&-{\sum_k}'(x_l-x_k)^{-1}\overline{\langle l| U[H,\lambda]U^\dagger \Omega^*
-\Omega U^*[H^*,\lambda]U^{\rm T}|k\rangle\langle k| U[H,\lambda]U^\dagger \Omega^*
-\Omega U^*[H^*,\lambda]U^{\rm T}|l\rangle}
\nonumber\\
  ={}&
  -2N\tau x_l
+2\tau q \delta_{l\bar l}
+2\tau(1-x_l)
-4\tau{\sum_k}'\frac{
x_lx_k-\delta_{l\bar l}\delta_{k\bar k}
-\delta_{k\bar l}(x_l+x_{\bar l})/2
+x_lx_k\delta_{lk}
 }{x_l-x_k},
\end{align}
\end{widetext}
where the $\delta_{lk}$ term can be dropped  because of the constraint
$k\neq l$ on the sum. Note how $Q$ changes to $q=|Q|$ because of the sign of
the matrix element involving pinned eigenvalues.

Again we find that eigenvalues at unity remain pinned.
For the other eigenvalues, we separate out from the sum the $q$ eigenvalues that are
pinned, and sum over the $M=(N-q)/2$ pairs of unpinned eigenvalues,
\begin{align}
\overline{\delta x_l}
={}&
  -2N\tau x_l
+2\tau(1-x_l)
-2\tau\frac{2-(x_l+x_{\bar l})}{
x_l-x_{\bar l}}
\nonumber
\\
&-4\tau q\frac{x_l
 }{x_l-1}
-4\tau x_l
{\sum_k}''
\frac{
x_k+x_{\bar k}-2x_{\bar l}
 }{x_l+x_{\bar l}-x_k-x_{\bar k}}
,
\end{align}
where the double-prime again indicates the exclusion of the pinned eigenvalues.
Furthermore,
\begin{eqnarray}
\overline{\delta x_l\delta x_m}&=&
8\tau(\delta_{l\bar m}-\delta_{l m}x_l^2).
\end{eqnarray}
For the quantities
 $\mu_l=(x_l+x_{\bar l})/2$, this gives
\begin{align}
&\overline{\delta\mu_l}
  =
  -2q\tau (\mu_l+1)
+2\tau(1-3\mu_l)
-4\tau
{\sum_k}''
\frac{
\mu_l^2-1
 }{\mu_l-\mu_k},
\\
&
\overline{\delta\mu_l\delta\mu_m}
=8\tau(1-\mu_l^2)\delta_{lm}.
\end{align}

The stationarity condition \eqref{stationary} is now fulfilled for
\begin{equation}
P(\mu_{1},\mu_{2},\ldots\mu_{M})\propto\prod_{k=1}^{M}(1-\mu_k)^{(q-1)/2}\,\prod_{l<m=1}^{M}|\mu_l-\mu_m|,\label{PmuBDI}
\end{equation}
which gives the
joint probability distribution
of the twofold degenerate, unpinned eigenvalues $\mu_{n}$ ($n=1,2,\ldots M$).

\subsection{Eigenvalue density}

The probability distributions \eqref{PmuD} and \eqref{PmuBDI} are both of the form
\begin{equation}
P(\mu_1,\mu_2,\ldots \mu_M)\propto\prod_{k=1}^{M}(1+\mu_k)^{a}(1-\mu_k)^{b}\,\prod_{l<m=1}^{M}|\mu_l-\mu_m|^{\beta},\label{PmuJacobi}
\end{equation}
with $\beta=2$, $a=1/2$, $b=|Q|-1/2$ in class D and $\beta=1$, $a=0$, $b=|Q|/2-1/2$ in class BDI. These are called Jacobi distributions, because the eigenvalue density $\rho(\mu)$ can be written in terms of Jacobi polynomials \cite{For10}.

For small $N$ it is quicker to calculate the eigenvalues density by integrating out all $\mu_{n}$'s except a single one. Keep in mind that $|Q|$ of the $\mu_{n}$'s are pinned at unity, and that the $N-|Q|=2M$ unpinned $\mu_{n}$'s are twofold degenerate. (The products in Eq.\ \eqref{PmuJacobi} run only over these $M$ unpinned pairs.) The eigenvalue density $\rho(\mu)=\langle\sum_{n=1}^{N}\delta(\mu-\mu_{n})\rangle$ is then given by
\begin{equation}
\begin{split}
\rho(\mu)={}&|Q|\delta(\mu-1)+2Mp(\mu),\\
p(\mu)={}&\int_{-1}^{1}d\mu_{1}\int_{-1}^{1}d\mu_{2}\\
&\cdots\int_{-1}^{1}d\mu_{M}\,\delta(\mu-\mu_{1})P(\mu_1,\mu_2,\ldots \mu_M).
\end{split}
\end{equation}
The delta functions satisfy $\int_{-1}^{1}\delta(\mu\pm 1)d\mu=1$. The average conductance follows from the eigenvalue density according to Eq.\ \eqref{Gav},
\begin{equation}
\langle G\rangle=T^{2}G_{0}\int_{-1}^{1}d\mu\,\rho(\mu)[1+(1-T)^2-2(1-T)\mu]^{-1}.\label{Gavmu}
\end{equation}
This gives the small-$N$ results in Eq.\ \eqref{GDsmallN} and Fig.\ \ref{fig_smallNpeak}. 

The large-$N$ limit \eqref{GDlargeN} is obtained from an integral equation for the eigenvalue density in the Jacobi ensemble \cite{Bee97,Dys72},
\begin{align}
&M\int_{-1}^{1}d\mu\,p(\mu')\ln|\mu-\mu'|=-\tfrac{1}{2}(1-2/\beta)\ln p(\mu)\nonumber\\
&\quad\quad-\frac{a}{\beta}\ln(1+\mu)-\frac{b}{\beta}\ln(1-\mu)+C+{\cal O}(1/M).
\end{align}
The constant $C$ is determined by the normalization
\begin{equation}
\int_{-1}^{1}d\mu\,p(\mu)=1.
\end{equation}
The solution is
\begin{align}
&Mp(\mu)=\frac{\tilde{M}}{\pi\sqrt{1-\mu^{2}}}-\frac{a}{\beta}\delta(\mu+1)-\frac{b}{\beta}\delta(\mu-1)\nonumber\\
&\quad+\tfrac{1}{4}(1-2/\beta)[\delta(\mu+1)+\delta(\mu-1)]+{\cal O}(1/M),\\
&\tilde{M}=M+(a+b)/\beta-\tfrac{1}{2}(1-2/\beta).
\end{align}

Upon substitution of the values for $a,b,\beta$ in the two symmetry classes, and transforming back from $p$ to $\rho$, we find
\begin{equation}
\rho(\mu)=\frac{N}{\pi}\frac{1}{\sqrt{1-\mu^{2}}}+\tfrac{1}{2}\delta(\mu-1)-\tfrac{1}{2}\delta(\mu+1)+{\cal O}(1/M),\label{rhomulargeN}
\end{equation}
independent of $Q$ and for both symmetry classes D and BDI. The corresponding result for the conductance is Eq.\ \eqref{GDlargeN}, to order $1/N$ if the limit $N\rightarrow\infty$ is taken at fixed $Q$.

\subsection{Large-voltage limit}

For completeness we also give the derivation of the large-voltage limit \eqref{Gtildeav} of the average conductance. We need to evaluate
\begin{equation}
\langle \tilde{G}\rangle=T^{2}G_{0}\int_{0}^{2\pi}d\phi\,\tilde{\rho}(\phi)\left|1-(1-T)e^{i\phi}\right|^{-2},\label{Gavapp}
\end{equation}
with $\tilde{\rho}(\phi)=\langle\sum_{n}\delta(\phi-\phi_{n})\rangle$ the density on the unit circle of the eigenvalues $e^{i\phi_{n}}$ of a unitary matrix $\tilde{\Omega}$. 

In class D the matrix $\tilde{\Omega}\equiv U$ is uniformly distributed in ${\cal U}(N)$. This is the circular unitary ensemble (CUE, $\beta=2$). In class BDI the chiral symmetry enforces that $\tilde{\Omega}$ is unitary symmetric, $\tilde{\Omega}=UU^{\rm T}$ with $U$ uniform in ${\cal U}(N)$. This is the circular orthogonal ensemble (COE, $\beta=1$). Unlike the probability distributions we needed for the zero-voltage limit, these two distributions are in the literature \cite{For10},
\begin{equation}
P(\phi_1,\phi_2,\ldots\phi_N)\propto\prod_{k<l=1}^{N}|e^{i\phi_{k}}-e^{i\phi_{l}}|^\beta.\label{Pphibeta}
\end{equation}

The corresponding density
\begin{equation}
\tilde{\rho}(\phi)=N/2\pi,\;\;0<\phi\leq 2\pi,
\end{equation}
 is uniform irrespective of the value of $\beta$ and without any finite-$N$ corrections. Substitution into Eq.\ \eqref{Gavapp} gives the result \eqref{Gtildeav}.
 
\section{Weak antilocalization in the circular real ensemble}
\label{WALcircular}

The results in Fig.\ \ref{fig_smallNpeak} for the zero-bias conductance peak are calculated in a random-matrix model where the electron and hole modes are mixed separately, but not together. Alternatively, we can consider what happens when all modes are uniformly mixed. This would be appropriate when the superconductor is connected to the tunnel barrier by a quantum dot, rather than by a disordered wire. In symmetry class D the reflection matrix $R$ at the Fermi level is then distributed according to the Poisson kernel of the circular real ensemble (CRE) \cite{Alt97,Ber09b}.

The calculation of the weak antilocalization peak proceeds as follows. The $2N\times 2N$ unitary reflection matrix $R$ determines the conductance according to
\begin{equation}
G/G_{0}=\tfrac{1}{2}N-\tfrac{1}{4}\,{\rm Tr}\,R\tau_{z}R^{\dagger}\tau_{z},\;\;\tau_{z}=\begin{pmatrix}
1&0\\
0&-1
\end{pmatrix},\label{Gtauzrelation}
\end{equation}
with $G_{0}=2e^2/h$. The Pauli matrix $\tau_z$ acts on the electron-hole degree of freedom. One readily checks, using unitarity of $R$, that this expression for the conductance is equivalent to Eq.\ \eqref{Gdef}. It is convenient to transform from the electron-hole basis to the Majorana basis,
\begin{equation}
R\mapsto{\cal U}R{\cal U}^{\dagger},\;\;{\cal U}=\sqrt{\frac{1}{2}}\begin{pmatrix}
1&1\\
-i&1
\end{pmatrix}.\label{RMajoranabasis}
\end{equation}
Since ${\cal U}\tau_{z}{\cal U}^\dagger=-\tau_{y}$, in the Majorana basis the conductance is given by
\begin{align}
G/G_{0}=\tfrac{1}{2}N-\tfrac{1}{4}\,{\rm Tr}\,R\tau_{y}R^{\dagger}\tau_{y}.\label{Gtauyrelation}
\end{align}

Particle-hole symmetry in the Majorana basis requires $R(-E)=R^{\ast}(E)$, where the excitation energy $E$ is measured relative to the Fermi level. At the Fermi level, $E=0$, this symmetry constrains $R$ to the group ${\rm O}(2N)$ of $2N\times 2N$ real orthogonal matrices. The topological quantum number $Q\in\{0,1\}$ is given by its determinant \cite{Akh11},
\begin{equation}
Q=\tfrac{1}{2}(1-{\rm Det}\,R),
\end{equation}
so that $R\in {\rm SO}(2N)\equiv {\rm O}_{+}(2N)$ in the topologically trivial system ($Q=0$, without Majoranas), and $R\in {\rm O}(2N)\backslash{\rm SO}(2N)\equiv {\rm O}_{-}(2N)$ in the topologically nontrivial system ($Q=1$, with Majoranas). This is the circular real ensemble (CRE). Away from the Fermi level, the constraint from particle-hole symmetry is ineffective and the reflection matrix ranges over the whole unitary group, $R\in {\rm U}(2N)$. This is the circular unitary ensemble (CUE).

The probability distribution of the scattering matrix $R_{0}$ without the tunnel barrier is uniform in ${\rm O}_{\pm}(2N)$ and ${\rm U}(2N)$ in the CRE and CUE, respectively: $P(R_{0})={\rm constant}$. The tunnel barrier, with a mode-independent transmission probability $T$, transforms $R_{0}$ into
\begin{equation}
R=\sqrt{1-T}+TR_{0}(1+\sqrt{1-T}\,R_{0})^{-1}.\label{RTR0}
\end{equation}
This introduces a nonuniformity in the probability distribution, described by the Poisson kernel \cite{Ber09b,Bro95b}
\begin{equation}
P(R)={\rm constant}\times|{\rm Det}\,(1-\sqrt{1-T}\,R)|^{-p}.
\label{PRPoisson}
\end{equation}
The exponent equals $p=4N$ in the CUE and $p=2N-1$ in the CRE.

\begin{figure}[tb]
\centerline{\includegraphics[width=0.9\linewidth]{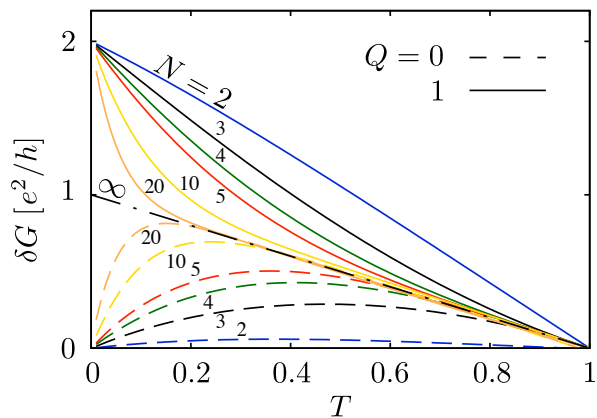}}
\caption{Amplitude $\delta G$ of the average zero-voltage conductance peak as a function of (mode-independent) transmission probability $T$, in symmetry class D for different number of modes $N$. The average is taken in the circular ensemble, either for a topologically trivial superconductor ($Q=0$, dashed curves) or for a nontrivial superconductor ($Q=1$, solid curves). The dash-dotted curve is the $Q$-independent large-$N$ limit \eqref{deltaGlargeNAZ}.
}
\label{fig_deltaGcircular}
\end{figure}

We have calculated the difference $\delta G=\langle G\rangle_{\rm CRE}-\langle G\rangle_{\rm CUE}$ from Eqs.\ \eqref{Gtauyrelation} and \eqref{RTR0}, upon averaging $R_{0}$ over ${\rm O}_{\pm}(2N)$ (for the CRE) and over ${\rm U}(2N)$ (for the CUE). [This numerical calculation was a quicker way to arrive at the answer than an analytical calculation using Eq.\ \eqref{PRPoisson}.] 

Results are shown in Fig.\ \ref{fig_deltaGcircular}. The large-$N$ limit has the $Q$-independent value \cite{Alt97}
\begin{equation}
\delta G=(1-T)\frac{e^{2}}{h}+{\cal O}(N^{-1}).\label{deltaGlargeNAZ}
\end{equation}
Comparison with Fig.\ \ref{fig_smallNpeak} shows that the two types of random-matrix models give qualitatively similar results.


\begin{thebibliography}{99}
\bibitem{note0} Chiral symmetry of the Hamiltonian is expressed by $H\mapsto -H$ upon exchange $e\leftrightarrow h$ of the electron and hole degrees of freedom. It is broken by spin-orbit coupling of transverse momentum, unlike the more fundamental particle-hole symmetry $H\mapsto -H^{\ast}$.
\bibitem{note2} For the parallel magnetic fields in Figs.\ \ref{fig_classD} and \ref{fig_Dsingle} we find that $Q=\frac{1}{2}(1-{\rm Det}\,r)$ switches from 0 to 1 at $E_{\rm Z}=8.1\,E_{\rm so}$ and then back to 0 at $E_{\rm Z}=11.3\,E_{\rm so}$, with a re-entrant $Q=0$ interval $9.2\,E_{\rm so}<E_{\rm Z}<9.7\,E_{\rm so}$.
\bibitem{paramfig3} The parameters for the simulations shown in Figs.\ \ref{fig_classD} and \ref{fig_Dsingle} are, in panel a): $x_{\rm barrier}=23\,l_{\rm so}$, $\delta L_{\rm barrier}=4a=0.4\,l_{\rm so}$, $U_{\rm disorder}=12.5\,E_{\rm so}$, $U_{\rm barrier}=15\,E_{\rm so}$, $G_{\rm disorder}=2.7\,e^{2}/h$, $G_{\rm barrier}=1.8\,e^{2}/h$; in panel b): $x_{\rm barrier}=3\,l_{\rm so}$, $\delta L_{\rm barrier}=a=0.1\,l_{\rm so}$, $U_{\rm disorder}=20\,E_{\rm so}$, $U_{\rm barrier}=50\,E_{\rm so}$, $G_{\rm disorder}=2.9\,e^{2}/h$, $G_{\rm barrier}=3.2\,e^{2}/h$.
\bibitem{note1} Fig.\  \ref{fig_classD} shows that the $Q=1$ peak becomes narrower than the $Q=0$ peak when the length of the disordered region between tunnel barrier and NS interface is reduced. This is consistent with the findings of Ref.\ \onlinecite{Pie12} for the effect of disorder on the Majorana resonance.
\bibitem{note3} The parameters for the simulations shown in Fig.\ \ref{fig_BDIav} are: $W=0.3\,l_{\rm so}$, $x_{\rm barrier}=2.7\,l_{\rm so}$, $\delta L_{\rm barrier}=a=0.01\,l_{\rm so}$, $E_{\rm F}=1000\,E_{\rm so}$, $\Delta_{0}=100\,E_{\rm so}$, $U_{\rm disorder}=4000\,E_{\rm so}$, $U_{\rm barrier}=7000\,E_{\rm so}$, corresponding to $N=8$, $G_{\rm disorder}=1.3\,e^{2}/h$, $G_{\rm barrier}=1.5 \,e^{2}/h$.
\end{thebibliography}
\end{document}